# Bright sub-20 nm cathodoluminescent nanoprobes for multicolor electron microscopy

Maxim B. Prigozhin*,[1], Peter C. Maurer*,[1], Alexandra M. Courtis[2], Nian Liu[3,4], Michael D. Wisser[3], Chris Siefe[3], Bining Tian[5], Emory Chan[5], Guosheng Song[6,7], Stefan Fischer[3], Shaul Aloni[5], D. Frank Ogletree[5], Edward S. Barnard[5], Lydia-Marie Joubert[8,9], Jianghong Rao[6], A. Paul Alivisatos[2,10,11,12], Roger M. Macfarlane[13], Bruce E. Cohen[5], Yi Cui[3], Jennifer A. Dionne[3], Steven Chu#,[1,14]

*These authors contributed equally to this work.

1. Department of Physics, Stanford University, Stanford, California 94305, United States

2. Department of Chemistry, University of California at Berkeley, Berkeley, California 94720, United States

3. Department of Materials Science and Engineering, Stanford University, Stanford, California 94305, United States

4. Current address: School of Chemical and Biomolecular Engineering, Georgia Institute of Technology, Atlanta, Georgia 30332, United States

5. Molecular Foundry, Lawrence Berkeley National Laboratory, Berkeley, California 94720, United States

6. Department of Radiology, Stanford University, Stanford, California 94305, United States

7. Current address: State Key Laboratory of Chemo/Biosensing & Chemometrics, College of Chemistry & Chemical Engineering, Hunan University, Changsha 410082, China

8. CSIF Beckman Center, Stanford University, Stanford, California 94305, United States

9. Current address: EM Unit, Central Analytical Facilities, Stellenbosch University, South Africa

10. Materials Sciences Division, Lawrence Berkeley National Laboratory, Berkeley, California 94720, United States

11. Department of Materials Science and Engineering, University of California, Berkeley, California 94720, United States

12. Kavli Energy NanoScience Institute, Berkeley, California 94720, United States

13. IBM Research-Almaden, San Jose, California 95120, United States

14. Department of Molecular and Cellular Physiology, Stanford University, Stanford, California 94305, United States

[#]e-mail: schu@stanford.edu

Electron microscopy (EM) has been instrumental in our understanding of biological systems ranging from subcellular structures[1] to complex organisms[2]. Although EM reveals cellular morphology with nanoscale resolution, it does not provide information on the location of proteins within a cellular context. An EM-based bioimaging technology capable of localizing individual proteins and resolving protein-protein interactions with respect to cellular ultrastructure would provide important insights into the molecular biology of a cell. Here, we report on the development of luminescent nanoprobes potentially suitable for labeling biomolecules in a multicolor EM modality. In this approach, the labels are based on lanthanide-doped nanoparticles[3] that emit light under electron excitation in a process known as cathodoluminescence (CL)[4]. Our results suggest that the optimization of nanoparticle composition, synthesis protocols and electron imaging conditions could enable high signal-to-noise localization of biomolecules with a sub-20-nm resolution, limited only **by the nanoparticle size. In ensemble measurements, these luminescent labels exhibit narrow spectra of nine distinct colors that are characteristic of the corresponding rare-earth dopant type.**

Nanoscale imaging of biomolecules in the context of cellular structures is essential to understand how cells function. Although conventional EM is a powerful tool for the study of heavy-metal-stained cellular ultrastructure[5] (*i.e.* lipid membranes, cytoskeleton, chromatin, *etc.*), it does not implicitly provide information about the location of specific biomolecules. Several approaches have been developed to overcome this limitation, most notably the tagging of target molecules with gold nanoparticles[6] and genetically encodable tags (*e.g.* recombinant ascorbate peroxidase[7,8]). However, these electron-contrast-based techniques are inherently limited to imaging one protein species at a time, which prevents studying protein-protein interactions and other complex processes. A related technology based on sequential photo-precipitation of small-molecule lanthanide ion complexes and subsequent multi-color imaging using the energy-filtered TEM has been reported[9]. However, this technology does not have single-molecule sensitivity and requires a distinctly addressable photosensitizer molecule for each lanthanide ion color, which limits the number of available spectroscopic channels. In contrast, when EM is combined with optical super-resolution microscopy, different proteins can be tagged with spectrally distinguishable labels[10]. Although promising, such correlative light and EM methods require challenging sample preparation, suffer from systematic errors due to sample disruption at the nanoscale[10], and are susceptible to background luminescence[11].

An alternative approach to visualizing multiple proteins in an electron micrograph relies on tagging proteins with fluorescent molecules or nanoparticles, which, under excitation by an electron beam, emit light in a process known as cathodoluminescence (CL). In principle, this method allows for a simultaneous acquisition of an electron micrograph and the locations of different proteins. However, organic dyes and fluorescent proteins rapidly disintegrate under electron exposure[12,13], and quantum dots are susceptible to bleaching in CL imaging[13]. Luminescent nanodiamonds and lanthanide-doped nanoparticles are more stable under electron beam irradiation and have been used for CL imaging, but only nanoparticles larger than ~40 nm have been reported to show detectable CL signal[14–16]. The large size of these nanoparticles prevents efficient protein labeling, which imposes a severe limitation on the use of these nanoparticles in biological experiments[17,18]. Here, we report on the development of cathodoluminescent lanthanide-doped $NaGdF_4$ and $NaYF_4$ nanocrystals with diameters less than 20 nm, which is comparable to quantum dots, gold nanoparticles, and immunoglobulin antibodies that are routinely used for immuno-labeling in electron microscopy[19,20].

In the work described here, a scanning electron microscope with a parabolic reflector is used to excite the cathodoluminescence of lanthanide-doped nanoparticles and image the CL signal onto a photo-multiplying detector (for a detailed description of the experimental setup see ref.[21] and the Methods section "Single-nanoparticle cathodoluminescence measurements"). In parallel with CL excitation and detection, the microscope also acquires the secondary electron (SE2) signal from the same pixels that are registered in the CL channel (see ref.[22] for software platform description). A schematic of the CL-SEM system used in this work (located at the Molecular Foundry of the Lawrence Berkeley National Laboratory) is shown in Fig. 1a. A representative transmission electron micrograph (TEM) of one of our nanoparticle samples ($NaGdF_4$:5% Eu) is shown in Fig. 1b. A key feature of cathodoluminescence imaging is its inherent nanoscale resolution. An electron beam with an energy of a few kiloelectron volts (keV) can be readily focused down to a few nanometers. However, in biological EM, the actual resolution is usually limited by sample preparation (*i.e.* fixation, sectioning, and heavy-metal staining[23]). In the case of CL imaging with nanoprobes, the resolution of protein localization will likely be limited by a combination of the CL excitation volume (see SI "Measurements and analysis of the electron beam sample interaction volume" and SI Fig. S13-15 for a discussion on CL excitation of rare-earth nanoparticles), nanoparticle size and nanoparticle surface functionalization.

As an example of the capability of cathodoluminescence microscopy, a CL-SEM image of a single $NaGdF_4$:5% $Eu^{3+}$ nanoparticle was acquired using both the secondary electron signal and the cathodoluminescence signal in parallel (Fig. 1c). Cross-sectional line profiles of the secondary electron (SE2, red) and CL (blue) signals from the same nanoparticle suggest that both EM and CL imaging can have comparable resolution typical of scanning electron microscopy. In contrast, a confocal optical scan of an upconverting lanthanide-doped nanoparticle ($NaYF_4$: 18% $Yb^{3+}$, 2% $Er^{3+}$) of a similar size shows a diffraction-limited point spread function typical of optical far field confocal microscopy. The specific nanoparticle shown in Fig. 1c was taken from a representative sample (see green star in Fig. 3b and a green data point in Fig. 3d). These data serve as an existence proof of the nanoscale resolution capability of CL-SEM for biological applications.

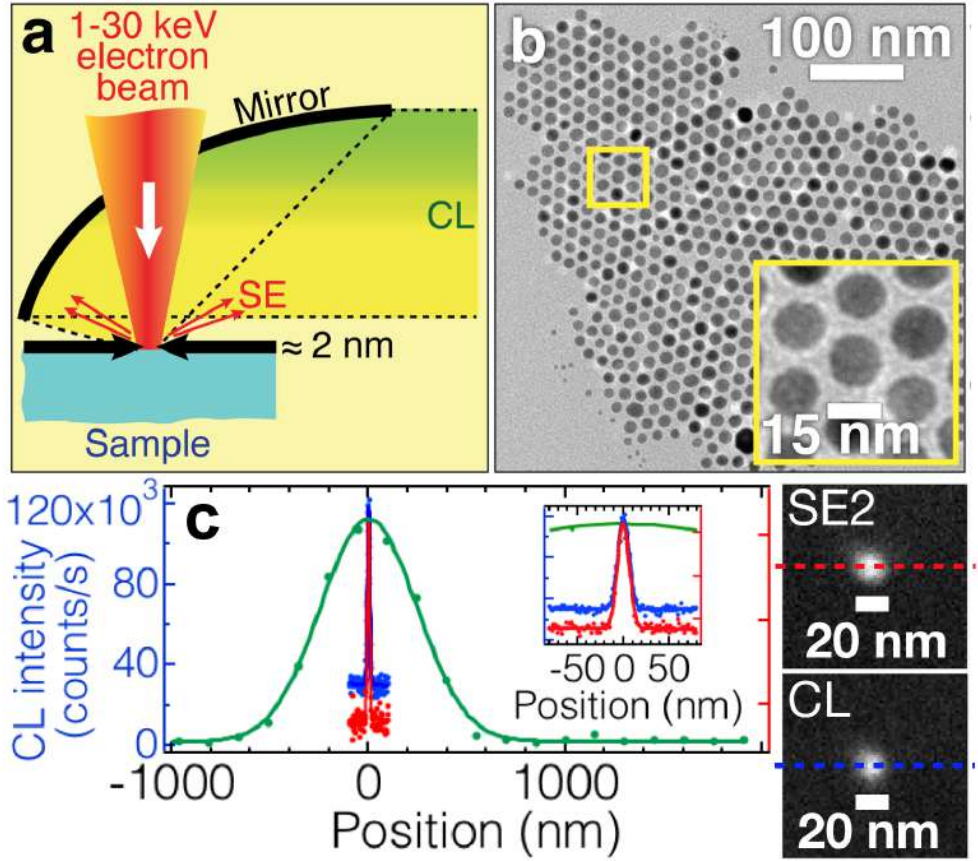

**Figure 1**. Cathodoluminescence microscopy concept. (a) Illustration of a cathodoluminescence instrument that uses an electron beam to induce emission of photons by nanoparticles shown in (b). CL emission is collected by a parabolic aluminum mirror and focused onto a photon-counting photomultiplier module. (b) TEM of $NaGdF_4$:5% $Eu^{3+}$ nanoparticles. Inset shows a magnified version of a region highlighted with a yellow square. (c) Simultaneous SEM and CL imaging of the particle highlighted with green stars in Fig. 3b. Cross-sectional line profiles of SE2 electrons (red, right axis) and CL (blue, left axis) scans of a single $BF_4^-$-exchanged $NaGdF_4$:5% $Eu^{3+}$ nanoparticle spin-coated on the Si substrate imaged in parallel. Pixel pitch is 1.95 nm; beam energy is 5 keV; pixel dwell time is 2 ms; beam current is ~400 pA. A cross-sectional line profile of a confocal light microscopy scan of a $NaYF_4$: 18% $Yb^{3+}$, 2% $Er^{3+}$ nanoparticle of similar size is shown in green (left axis). Excitation wavelength is 980 nm; water immersion objective with 1.27 numerical aperture was used. Inset shows the same data but with zoomed-

in *x*-axis. Raw SE2 and CL data are shown on the right. All cross-sectional line profiles are fit to a single-Gaussian model.

Previous work on lanthanide-doped nanoparticle synthesis has mainly been focused on obtaining nanoparticle compositions that optimize excitation and emission efficiency in optical upconversion[24]. However, the excitation mechanism for CL in the energy range used in these studies is fundamentally different from that of upconversion[4]. When lanthanide-doped nanoparticles are used for upconversion or ordinary fluorescence, they are normally illuminated by a monochromatic light source, and the multicomponent emission spectrum is a result of energy transfer among different color centers and between excited states[25]. In CL, the incident high energy electrons excite valence band electrons of the nanocrystal into a broad continuum of excited conduction band states and phonon-broadened excited impurity states. The high energy electron states cascade to lower energies that include photoluminescent states. Therefore, an independent investigation of the CL brightness of rare-earth nanoparticles as a function of their composition was required. Such optimization was achieved by synthesizing a series of $NaGdF_4$ and $NaYF_4$ nanoparticles of varying $Eu^{3+}$ doping levels (see the Methods section "Nanoparticle synthesis and characterization") and characterizing their CL brightness at the single-nanoparticle level (See Methods). Nanoparticles were synthesized using a colloidal synthesis method as described in the Methods[26,27,28,29]. The as-synthesized nanoparticles were ligand-exchanged with nitrosonium tetrafluoroborate ($NOBF_4$) and dispersed in dimethylformamide (DMF) (see Methods section "Sample preparation for single-nanoparticle cathodoluminescence measurements")[30]. For CL-SEM imaging, multiple samples from the same synthesis run were prepared by spin-coating the dispersion of nanoparticles in DMF on a silicon substrate. The concentration of the nanoparticle solution used for spin-coating was adjusted so that at least three isolated nanoparticles could be found in a given field of view with an area of 1 μm$^2$. The samples were imaged in CL-SEM as described in the Methods ("Single-nanoparticle cathodoluminescence measurements" section).

Figure 2a-b shows a sample CL-SEM image of NaGdF$_4$:5% Eu$^{3+}$ nanoparticles. This sample is indicated by the magenta arrow in Fig. 3b and a magenta data point in Fig. 3d. In a typical experiment, a 1 μm$^2$ field of view was imaged with a 1.95 nm pixel pitch (comparable to the typical electron beam size) and a pixel dwell time of 2 ms using a 5 keV electron beam with a current of ~400 pA (current density ~100 pA/nm$^2$; ~6.2 x 10$^6$ electrons s$^{-1}$ Å$^{-2}$; dose is ~12.5 x 10$^3$ electrons Å$^{-2}$ within the 2 ms pixel dwell time). The secondary electron (SE2) image (Fig. 2a) was collected in parallel with the CL image (Fig. 2b). In order to extract the CL intensity and signal-to-noise ratio (SNR) for individual nanoparticles, a sub-region within the original 1 μm$^2$ field of view containing one or several individual nanoparticles was selected. Nanoparticle aggregates were deliberately excluded from the analysis. For the case when a single nanoparticle was selected, its raw image was fitted to a two-dimensional Gaussian function with a linearly sloped background. This fit was used to extract the CL intensity and the signal-to-noise ratio for each individual nanoparticle (See Methods section "Single-nanoparticle cathodoluminescence data analysis"). The pixel pitch of 1.95 nm is significantly smaller than the SEM nanoparticle full-width-half-maximum (FWHM) of ~20 nm, leading to oversampling. An improvement in signal quality can be achieved through Fourier filtering. Fig. 2c-e shows the CL intensity after Gaussian-filtering the image. In Fig. 2d, the CL counts per second and the SNR of all the individual nanoparticles seen in images in Fig. 2a,b are shown. CL intensity data are fit to a cubic curve for comparison with volumetric scaling of CL intensity, and SNR data are fit to a power law with an exponent of 3/2 due to the square root scaling of the noise with respect to the signal. Fig. 2f,g depict three-dimensional visualizations of the CL signal of individual nanoparticles highlighted with red and yellow boxes in Fig. 2c, respectively. The corresponding SEM scan in Fig. 2a proves that the CL emission originates from individual nanoparticles. For example, the dimmer particle in the red rectangle in Fig. 2c has a SNR of 15.7 and is clearly visible in the filtered CL image (Fig. 2f).

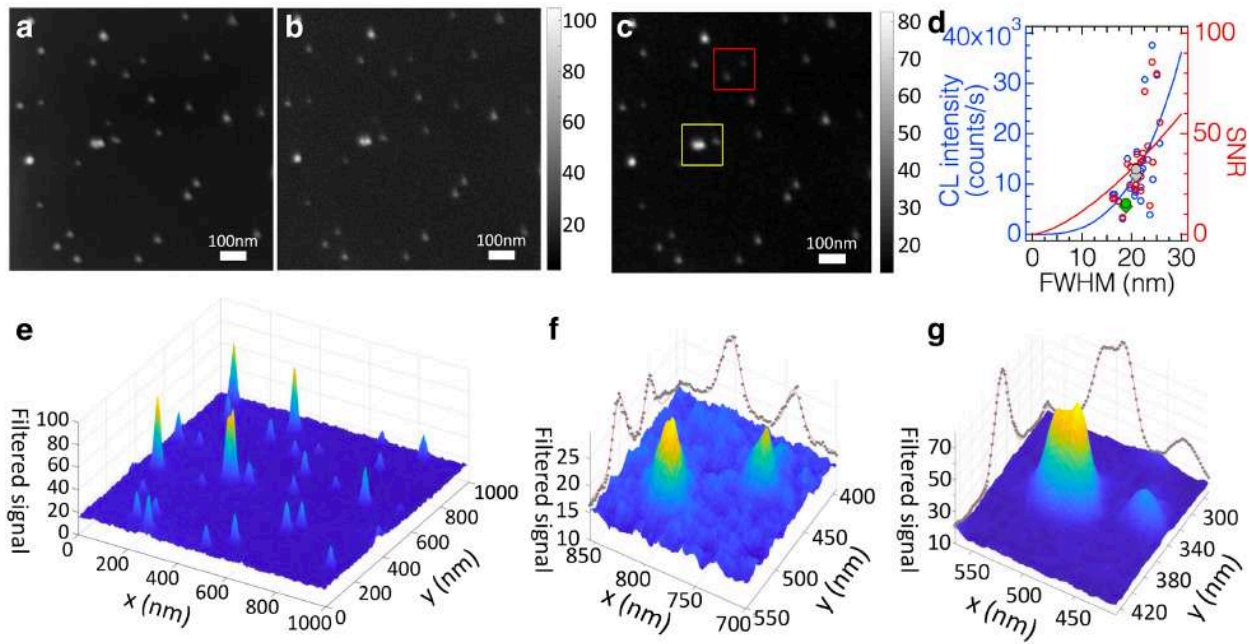

**Figure 2.** Cathodoluminescence imaging of single $BF_4^-$-exchanged $NaGdF_4$:5% $Eu^{3+}$ nanoparticles spin-coated on the Si substrate. (a) Raw unfiltered secondary electron (SE2) image and (b) CL signal (detected photons) of $NaGdF_4$:5% $Eu^{3+}$ nanoparticles (highlighted with a magenta arrow in Fig. 3b and a magenta data point in Fig. 3d) collected in parallel. Images are 1 $\mu m^2$; pixel pitch is 1.95 nm; beam energy is 5 keV; pixel dwell time is 2 ms (image acquisition time is 8 min 44 s); beam current is ~400 pA. (c) Fourier-filtered CL image using a Gaussian smoothing function of $\sigma$ = 7.16 nm. The smoothed red and yellow regions in panel (c) are shown in panels (f) and (g), respectively. (d) CL intensity (left axis, blue) and the signal-to-noise ratio (SNR; right axis, red) for all the single nanoparticles in panels (a-b) plotted as a function of the full-width-half-maximum (FWHM, 2.35$\sigma$) of the two-dimensional Gaussian

distribution fitted to each nanoparticle CL signal. A fit of volume scaling of CL intensity (FWHM$^3$, blue curve, left axis) and a corresponding FWHM$^{3/2}$ fit for the SNR (red curve, right axis) are shown. The data points for the nanoparticles in panel (f) are highlighted with diamonds (CL signal, left axis) and circles (SNR, right axis). The brighter nanoparticle from panel (f) is in gray, the dimmer nanoparticle is in light green. (e) Three-dimensional smoothed plot of the CL data shown in panel (c). (f, g) Images generated using the same filtering conditions as in panels (c, e) but for the zoomed in red and yellow regions in panel (c). A doublet of neighboring nanoparticles is shown in panel (g). The dotted points on the sides correspond to a maximum intensity projection of the filtered CL signal, and the red solid lines represent a fitted Gaussian as a guide to the eye.

Despite high CL intensity in certain samples, the CL emission of NaGdF$_4$:Eu$^{3+}$ nanoparticles varied drastically between different experiments. For example, particles that were synthesized under nominally identical conditions resulted in dramatically different CL emission rates (Fig. 3c). Similar sample preparation resulted in a large variation in signal, even if these samples were prepared from the same nanoparticle stock solution (Fig. 3b). Finally, even on a single sample (*i.e.* a single 5 x 5 mm Si wafer substrate), the CL signal fluctuated as a function of the imaging position (Fig. 3a, and SI Figs. S16-S22). In the field of view (1 μm$^2$) that contains the nanoparticles with the largest CL emission rate (Position 5 in Fig. 3a, SI Fig. S20) an average CL signal of (7.1 ± 2.2) x 10$^5$ counts/s was observed, but in other cases (*e.g.* Position 7 in Fig. 3a, SI Fig. S22) the nanoparticles imaged by the CL-SEM were barely detectable in the CL channel. The high variability of the CL signal for NaGdF$_4$:Eu$^{3+}$ nanoparticles may originate from material-specific or synthesis-specific defects that lead to quenching of the luminescence. Such quenching can be caused, for example, by electron beam damage of the surrounding organic material or the nanoparticle itself[31]. CL brightness of NaGdF$_4$ nanoparticles as a function of Eu$^{3+}$ doping level was also investigated (Fig. 3d). NaGdF$_4$ doped with Er$^{3+}$ at various doping levels showed no detectable CL luminescence at the single-nanoparticle level (Fig. S23a). Notably, the large variability in the CL brightness of individual NaGdF$_4$:Eu$^{3+}$ nanoparticles was masked in ensemble CL measurements (Fig. S25). This observation highlights the importance of single-nanoparticle CL measurements. In addition, CL brightness of ~35 nm FWHM NaYF$_4$ nanoparticles doped with Eu$^{3+}$ (Fig. S24) and with Er$^{3+}$ (Fig. S23b) was more consistent from sample to sample as compared to the NaGdF$_4$ nanoparticles. Further CL nanoprobe development is needed to achieve sub-20-nm nanoparticles with consistently high CL luminescence required for biological cell imaging.

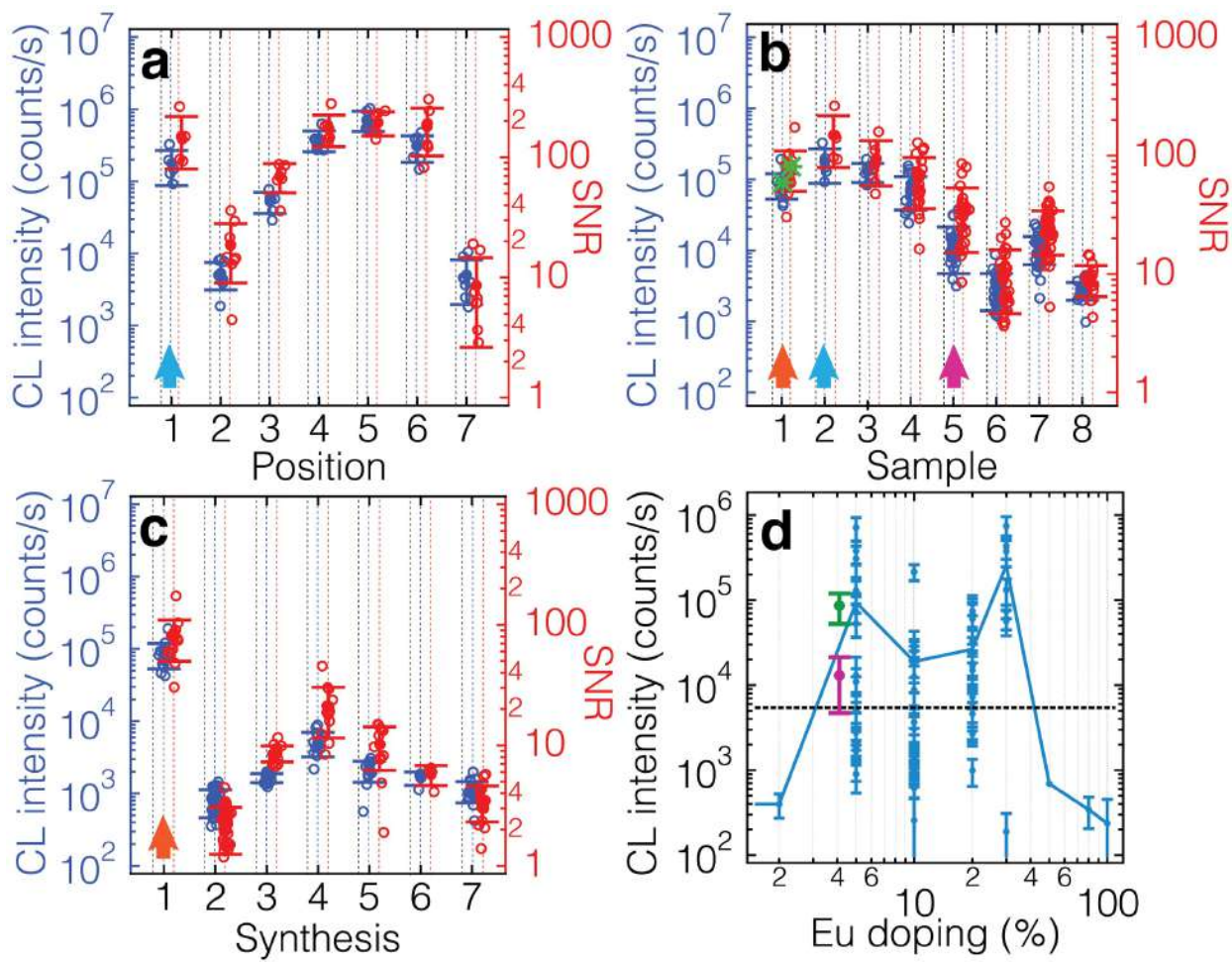

**Figure 3.** Variability of CL brightness and signal-to-noise ratio for BF$_4^-$-exchanged Eu$^{3+}$-doped NaGdF$_4$

nanoparticles spin-coated on the Si substrate (a) within the same sample (Sample #2 in panel (b)) of the brightest synthesis run (Synthesis #1 in panel (c)), (b) within synthesis #1 but for different samples (*i.e.* separately prepared Si substrates), and (c) across different synthesis runs. Identical samples are shown with two light blue arrows in panels (a-b) and two orange arrows in panels (b-c). Data used for the single nanoparticle in Fig. 1c is highlighted with green stars in panel (b) and the data set for the corresponding sample region is highlighted with a green data point in panel (d). A region of interest shown in Fig. 2 is highlighted with a magenta arrow in panel (b) and a magenta data point in panel (d). Each blue empty circle corresponds to a single nanoparticle within a 1 μm$^2$ field of view and represents the brightness (Gaussian amplitude) in counts per second (blue, y-axis on the left). Each red empty circle represents the signal-to-noise ratio for one nanoparticle. Filled circles represent the average brightness in counts per second (blue, y-axis on the left) and the signal-to-noise ratio (SNR, red, y-axis on the right). Error bars represent one standard deviation from the mean. (d) Single nanoparticle measurements of CL brightness as a function of Eu$^{3+}$ doping concentration for the NaGdF$_4$ host. Each data point corresponds to an average of nanoparticle brightness in a single 1 μm$^2$ field of view and the corresponding error bars show one standard deviation. Green and magenta data points represent data sets used in Fig. 1c and Fig. 2, respectively. These data points also correspond to 5% Eu$^{3+}$ doping level, but the data points are offset to the left for visibility. All data are plotted including images from different syntheses, different samples from the same synthesis, and data from different regions of the same sample. The solid blue line represents an average of the intensities of all measurements at each individual Eu$^{3+}$ doping level. The black horizontal dashed line in panel (d) represents the average noise level for data shown in panels (a-c) (5449 counts/s) to illustrate the level of marginal CL intensity.

Lanthanide ions have rich energy level diagrams with 4*f*-to-4*f* transitions that give rise to emission spectra that are characteristic of each individual lanthanide ion. Fig. 4 shows ensemble CL spectra obtained for nine different types of NaGdF$_4$ nanoparticles doped with Eu$^{3+}$, Er$^{3+}$, Ho$^{3+}$, Tb$^{3+}$, Sm$^{3+}$, Dy$^{3+}$, Nd$^{3+}$, Tm$^{3+}$ and Yb$^{3+}$ ions (see Fig. S6 for TEM images of NaYF$_4$ nanoparticles and Fig. S26 for the CL spectra of NaYF$_4$ nanoparticles). For the NaGdF$_4$ host, Er$^{3+}$ was doped at 20 %, while all other lanthanide ions were doped at 15 %. All the spectra were acquired from films of *n*-hexane-washed nanoparticles using a JEOL JXA-8230 SuperProbe instrument (beam energy is 5 keV; probe current ~0.13 pA/nm$^2$, ~8 x 10$^3$ electrons s$^{-1}$ Å$^{-2}$; see Methods section "Ensemble cathodoluminescence measurements"). In addition, doping with Ce$^{3+}$ and Pr$^{3+}$, and Gd$^{3+}$ alone was investigated but did not yield sharp spectra (see SI Fig. S29 for the spectra of Ce$^{3+}$, Pr$^{3+}$, and Gd$^{3+}$).

The narrow emission lines (21 ± 11 nm FWHM) of lanthanide-doped nanoparticles and their invariance with respect to the host lattice are indicative of atom-like 4*f*-to-4*f* inner-shell transitions in lanthanide ions. The intensities of transitions among these low-lying 4*f* states in lanthanide ions can be qualitatively described by the Judd–Ofelt theory[32,33]. Fig. 4 compares the experimental data to the Judd–Ofelt theory (see SI section "Simulations of nanoparticle spectra", Fig. S30). Although these calculations are in good qualitative agreement with the obtained data, differences arise with respect to the relative oscillator strengths of individual transitions. These discrepancies likely arise because the Judd-Ofelt parameters rely on the rates extracted from different host lattices[34].

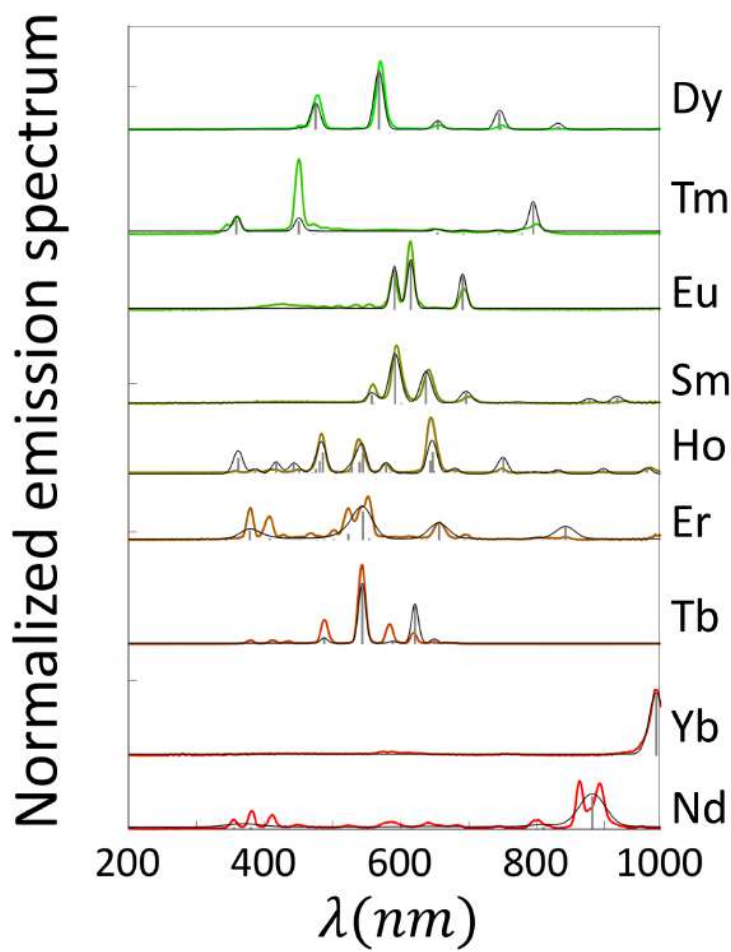

**Figure 4**. Multicolor imaging of lanthanide-based nanoprobes. (a) Emission spectra of ensemble samples of *n*-hexane-washed $NaGdF_4$ (solid colored lines) nanoparticles doped with different lanthanide ions (20 % doping for $Er^{3+}$; 15 % doping for the others) drop-cast on a Si substrate. The data were acquired at 5 keV electron beam energy, and corrected for spectrometer efficiency. Gray bars correspond to the Judd-Ofelt calculations (see Methods section "Simulations of nanoparticle spectra" and Fig. S30). The gray lines correspond to fitted spectra using linewidth as the only fitting parameter.

This work explores lanthanide-doped nanoparticles as prospective nanoscale labels for multicolor electron microscopy owing to their potentially high photon count rate, sharp emission spectra, and tunable size. The size of lanthanide-doped nanoparticles presented here is comparable to that of quantum dots that are commonly used as luminescent labels in optical imaging[20], and only slightly larger than gold nanoparticles used in electron microscopy[19]. A further reduction in size may be achieved by engineering core-shell structures that eliminate the adverse effects of the nanoparticle surface[35,36]. Furthermore, the long excited state lifetimes of rare-earth nanoparticles[37] enable time-gated measurements that eliminate CL background from the biological substrate[38] (Fig. S28) and would allow imaging even smaller nanoparticles that have a lower CL intensity but are more suitable for targeted protein labeling and penetration into tissue samples[18]. Although the focus of the work presented here is on the development of bright cathodoluminescent nanoparticles, the CL background from potential biological substrates needs to be taken into account when designing optimal CL nanoprobes. Background-free measurements would make the detection of a few photons sufficient to successfully assign the nanoparticle color (see SI section "Estimation of number of observable colors and required photon count rate"), potentially opening a path to ultra-small labels for multicolor biological electron microscopy.

A better signal-to-noise ratio for the CL of rare-earth nanoprobes may be achieved if the electron interaction volume is matched to the size of the nanoparticle (See SI section "Measurements and analysis of the electron beam sample interaction volume"). Theoretical analysis of inelastic scattering suggests that for a 15-20 nm diameter nanoparticle, the electron interaction volume would match the nanoparticle dimensions at the electron landing energy of 0.75-1 keV (Fig. S15). In addition to allowing more efficient energy deposition into a nanoparticle, the local excitation provided by the low electron landing energy minimizes the background by reducing the CL "halo" that originates from the excitation of the nanoparticles by back-scattered and secondary electrons in the substrate (Fig. S13-14). Landing electron energy of ~1 keV is also an optimal trade-off between the back-scattered electron contrast and the axial resolution in biological SEM[39]. Furthermore, because the ions that comprise rare-earth nanoparticles have high atomic numbers compared to the constituents of the biological tissue, a positive identification of a nanoparticle can be done using the secondary or back-scattered electron signal. Once a nanoparticle is identified in the electron imaging channel, a sufficient number of CL counts is only required to identify the spectral identity of this nanoparticle, which slightly relaxes the constraints on the required signal-to-noise ratio (See Methods, SI section "Single-particle signal-to-noise ratio" and Fig. S12, as well as the SI section "Estimation of the number of observable colors and the required photon count rate"). In the future, rapid nanoparticle localization in the electron detection channel may allow addressing each identified nanoparticle individually with an electron beam to determine its color (*i.e.* spectral fingerprint) in the CL channel instead of scanning the entire field of view with the slow 2 ms pixel dwell time that was required for CL imaging in this work. This new data acquisition strategy is expected to substantially decrease both the imaging time and the electron dose. Combining CL microscopy with the new multi-beam SEMs is expected to further increase the imaging speed.

Finally, the next generation of our experiment will focus on the optimization of the synthesis parameters for the other lanthanide ions (*i.e.* $Ho^{3+}$, $Tb^{3+}$, $Sm^{3+}$, $Dy^{3+}$, $Nd^{3+}$, $Tm^{3+}$ and $Yb^{3+}$), similar to that done for $Eu^{3+}$. Such an optimization, combined with a new multicolor-CL imaging system, which enables a simultaneous detection of multiple spectral components, may open the door to true multicolor imaging at the single-nanoparticle level. The potentially large photon count rate of individual nanoparticles (i.e. $NaGdF_4$:5% $Eu^{3+}$) combined with the distinct spectra obtained in ensemble measurements suggest that up to nine different colors with 10-20 nm spatial resolution could potentially be achieved (see SI section "Estimation of the number of observable colors and the required photon count rate", Fig. S27).[40] A further increase in the number of colors may be achieved by co-doping nanoparticles with multiple lanthanide ions and detecting an emission spectrum characteristic of the exact particle composition[34]. Incidentally, another benefit of using low beam energy in CL imaging lies in the imaging of several nanoprobes of different colors because the spectral identification would be compromised if stray electrons were able to excite neighboring nanoparticles (see SI section "Measurements and analysis of the electron beam sample interaction volume").

Although reliable multicolor CL imaging at the single-nanoparticle level and in biological tissue remains to be demonstrated, our findings motivate future work in this direction. Optimal multicolor imaging combined with advances in particle functionalization and labeling[41] could allow visualizing the locations of different proteins with respect to the cellular ultrastructure (*i.e.* organelles, vesicles, nucleic acids, and other nanostructures). Protein-specific localization in the context of ~ 5 nanometer cell ultrastructure could have a significant impact on our understanding of the molecular architecture of the cell. Likewise, combining multicolor CL imaging with recent advances in *in situ* serial-block-face SEM[23] or focused ion beam (FIB) SEM[42] will permit full three-dimensional reconstruction of the entire tissue sections[43,44], while providing simultaneous nanoscale protein localization. Such bio-specific volumetric electron imaging would enable the visualization of different cell types within heterogeneous tissue sections and shed light onto the organization of complex systems such as the heart[45], the brain[43], or cancerous tissue[46].

**Acknowledgements**

This work was supported by the Gordon and Betty Moore Foundation grant number 4309. Work at the Molecular Foundry was supported by the Office of Science, Office of Basic

Energy Sciences, of the U.S. Department of Energy under Contract No. DE-AC02-05CH11231. Part of this work was performed at the Stanford Nano Shared Facilities (SNSF), supported by the National Science Foundation under award ECCS-1542152. Access to the JEOL TEM 1400 was provided through the Stanford Microscopy Facility, NIH grant SIG number 1S10RR02678001. M.B.P. was supported by the Helen Hay Whitney Foundation Postdoctoral Fellowship, and P.C.M. was supported through the Stanford Neuroscience Interdisciplinary Award. M.D.W. and J.A.D. acknowledge financial support provided as part of the DOE "Light-Material Interactions in Energy Conversion" Energy Frontier Research Center under grant DE-SC0001293, as well as funding provided by the Global Climate and Energy Project at Stanford University. The authors thank Prof. Naomi Ginsberg and Dr. Clarice Aiello for providing software used in time-gated cathodoluminescence imaging experiments and for stimulating discussions. The authors are also grateful to Dr. John J. Perrino and Dr. Dale H. Burns for providing expertise in electron microscopy and for stimulating discussions. We thank Prof. Ronald Walsworth, Dr. David Glenn, and Dr. Huiliang Zhang; Prof. Thomas C. Sudhof and Dr. Justin Trotter; Dr. Joshua Collins; and Dr. Xianchuang Zheng for stimulating discussions.

**Methods**

**Nanoparticle synthesis and characterization.** A series of nanoparticles was synthesized including $NaGdF_4$:$Eu^{3+}$ and $NaGdF_4$:$Er^{3+}$ doped at 0, 2, 5, 10, 20, 30, 50, 80, and 100 %; $NaYF_4$:$Eu^{3+}$ doped at 2, 5, 10, and 20 %; $NaYF_4$:$Er^{3+}$ doped at 5, 10, and 30 %; $NaGdF_4$ nanoparticles doped with $Ho^{3+}$, $Tb^{3+}$, $Sm^{3+}$, $Dy^{3+}$, $Nd^{3+}$, $Tm^{3+}$, $Ce^{3+}$, $Pr^{3+}$, and $Yb^{3+}$ at 15 %; and $NaYF_4$ nanoparticles doped with $Ho^{3+}$, $Tb^{3+}$, $Sm^{3+}$, $Dy^{3+}$, $Nd^{3+}$, $Tm^{3+}$, $Ce^{3+}$, $Pr^{3+}$, and $Yb^{3+}$ at 5 %. Synthesis was based on the previously reported protocols[26–29] (see SI section "Nanoparticle synthesis and characterization" for details). Briefly, 4 mL of oleic acid and 6 mL of 1-octadecene were mixed with 0.4 mL total volume (0.4 mmol) of an aqueous solution of 1 M rare-earth (RE) chloride hydrates of desired ratios. The temperature was set to 110 °C for 40 minutes. Afterwards, the solution was cooled to < 30 °C. Next, a nucleation precursor solution was prepared by adding 1 mL of 1 M sodium hydroxide in methanol to 4 mL of 0.4 M ammonium fluoride in methanol. After mixing, the precursor solution was vortexed for 10 s and injected into the RE-oleate mixture at room temperature under argon atmosphere. The temperature was maintained at 50 °C under argon atmosphere for 40 minutes. The temperature was further increased to 80 °C and the reaction was either exposed to air or put under vacuum allowing methanol to evaporate. The temperature was stabilized at 100 °C for 15 min under vacuum. Afterwards, the reaction was placed under argon atmosphere and the temperature was increased to 300 °C ($NaYF_4$) or 320 °C ($NaGdF_4$) at a mean rate of ~15 °C/min. The growth temperature was maintained for 60 ($NaGdF_4$) or 90 ($NaYF_4$) minutes before cooling the reaction to < 30 °C. The samples were stored as-synthesized in oleic acid 1-octadecene. Nanoparticle characterization was done using TEM.

**Sample preparation for single-nanoparticle cathodoluminescence measurements.** The nanoparticles were exchanged into dimethylformamide (DMF) using a modification of a published procedure[30] (see SI section "Single-particle CL sample preparation" for details). Briefly, 0.5-1 mL of as-synthesized nanoparticles were mixed with an equal volume of ethanol and washed by centrifugation at 3,500 g for 3 minutes. The pellet was resuspended with 0.5 mL *n*-hexane and the 0.5 mL ethanol wash was repeated. The pellet was then resuspended in 0.3 mL of *n*-hexane, 0.3 mL of 11 mg/mL nitrosonium tetrafluoroborate ($NOBF_4$) in DMF was added, and the reaction was incubated for 45 minutes. The tube was centrifuged at 10,000 g for 10 minutes and the supernatant was discarded. The pellet was washed with 0.2 mL of a 1:1 mixture of toluene and *n*-hexane at 10,000 g for 10 minutes. The resulting nanoparticle pellet was dried under argon and resuspended in 0.1-0.2 mL of DMF. 4 µL of nanoparticles in DMF were spin-coated on a silicon substrate to achieve a density compatible with imaging several single nanoparticles within a 1 µm² region.

**Single-nanoparticle cathodoluminescence measurements.** Single-nanoparticle CL measurements were done at the Molecular Foundry at the Lawrence Berkeley National Laboratory. Experiments were performed on a Zeiss Supra 55-VP-FESEM with a cathodoluminescence parabolic mirror light collection system. A 1.3π sr (1 mm focal length) diamond-turned aluminum parabolic reflector mounted on a 4-axis nanopositioning stage was used to collimate the light emitted from the sample. The light was then focused onto a photomultiplier counting module (Hamamatsu H7442-40). During the measurements, the working distance was typically in the 4.9-5.2 mm range, the current was on the order of 300-500 pA, and the beam energy was 5 keV. The samples were scanned using a 512 x 512 point grid of 1 µm x 1 µm dimensions resulting in a pixel pitch of 1.95 nm. The dwell time per pixel was either 500 µs or 2 ms depending on the experiment and expected count rates. The estimated electron dose under these conditions was approximately 5,000-20,000 electrons/Å² (current density ~100 pA/nm²).

**Single-nanoparticle cathodoluminescence data analysis.** The CL intensity and signal-to-noise ratio for individual nanoparticles is extracted by selecting a sub-region (approx. 30 x 30 pixels, or ~7-8 σ for NaGdF$_4$:Eu$^{3+}$) within the original 1 µ**m**² field of view, which contains one or several individual particles. Nanoparticle aggregates were avoided. The raw CL image $I(x_i, y_i)$, where $x_i$ and $y_i$ are discrete pixels of 1.95 nm pitch and *I* is measured in counts per pixel, was then fitted by a two-dimensional Gaussian function with a linearly sloped background of the form $G(x_i, y_i) = c_o + c_1 y_i + c_2 x_i + A e^{\frac{-(x_i-x_o)^2 - (y_i-y_o)^2}{2\sigma^2}}$ (Fig. S11). In the fit routine the starting parameters for the standard deviation σ and the center $x_0$ and $y_0$ positions were obtained from similar fits to the SE2 signal. The standard deviation of the CL image was constrained to not deviate by more than 10 % from the SE2 image. The Gaussian component of the fitted function $S(x_i, y_i) = A e^{\frac{-(x_i-x_o)^2 - (y_i-y_o)^2}{2\sigma^2}}$ represents the CL signal of an individual nanoparticle. The sum of $S(x_i, y_i)$ corresponds to the number of counts associated with each nanoparticle. Note, since the "counts" are derived from a fitted function, the sum is generally not an integer. The signal-to-noise ratio was calculated by first summing up the CL signal of the nanoparticle $S_{total} = \sum_i S(x_i, y_i)$ including only the pixels in which the signal is greater than a defined threshold, which is commonly taken as the signal at pixels $2\sigma$ (95% confidence level) away from the center of the Gaussian distribution. The threshold of $2\sigma$ was used because it is ideal when the background level is equal to the amplitude of the Gaussian of the fitted CL signal, which is the case for the data collected in this work. The noise was calculated for the same pixels as $N_i = \sqrt{G(x_i, y_i)}$ and the total noise was determined by adding the noise for each pixel in quadrature: $N_{total} = \sqrt{\sum_i N_i^2}$. The overall signal-to-noise ratio for a single nanoparticle was then defined as the ratio $SNR = \frac{S_{total}}{N_{total}}$. The full-width-half-maximum (FWHM) was calculated from the standard deviation, $\sigma$, as FWHM $= 2\sigma\sqrt{2\ln 2} = 2.35\sigma$. Note that in this analysis the CL signal is approximated by a Gaussian function, and does not include such imaging artifacts as astigmatism, sample drift, or charging.

**Ensemble cathodoluminescence measurements.** 0.5 mL of as-synthesized nanoparticle solution in oleic acid and 1-octadecene was washed three times with 0.5 mL ethanol at 3.5 g for 3 minutes at room temperature. The nanoparticles were redispersed in *n*-hexane and drop-cast repeatedly on a ~5 mm x 5 mm piece of silicon wafer until an opaque white film of nanoparticles was visible by eye. Nanoparticle spectra were measured with a JEOL JXA-8230 SuperProbe electron microscope equipped with an xCLent III hyperspectral cathodoluminescence system. See SI section "Ensemble spectral measurements and sample preparation" for further details.

**Simulations of nanoparticle spectra.** The luminescence spectra for different dopants ($Eu^{3+}$, $Er^{3+}$, $Ho^{3+}$, $Tb^{3+}$, $Sm^{3+}$, $Dy^{3+}$, $Nd^{3+}$, $Tm^{3+}$ and $Yb^{3+}$) were qualitatively estimated by the Judd-Ofelt theory[34]. A rate-equation-based model, which incorporates electric and magnetic dipole transitions, cross relaxations between multiple rare-earth ions, and multi-phonon relaxations in the host lattice was used (for more details on the software package we refer to Chan *et al.*[34]). Although electron excitation generally involves high-lying energy states, which cannot be described by the Judd-Ofelt theory, emission in the visible spectrum can be accurately modeled. The present Judd-Ofelt simulations were restricted to energy levels below 25,000 cm$^{-1}$.

**Data availability.** The datasets generated during and/or analyzed during the current study are available from the corresponding author on reasonable request.

**Author contributions**

M.B.P., P.C.M., and S.C. conceived the project, designed experiments, analyzed the data, and interpreted the results. M.B.P. and P.C.M. conducted cathodoluminescence imaging experiments and wrote software for data analysis. M.B.P., P.C.M., A.M.C., N.L., M.D.W., C.S., B.T., G.S., and S.F. synthesized and characterized rare-earth nanoparticles. E.C. provided software for the simulation of the nanoparticle spectra. S.A., D.F.O., E.B., and L.-M.J. provided assistance and expertise in electron microscopy hardware and sample preparation. D.F.O. and S.A. developed the CL optics, and E.S.B. and D.F.O. developed the CL software. S.C. supervised research. J.R., A.P.A., R.M.M., B.E.C., Y.C., and J.A.D. supervised the relevant portions of the research such as sample preparation and training M.B.P. and P.C.M. on the nanoparticle synthesis. M.B.P., P.C.M., and S.C. wrote the manuscript.

**Competing interests statement**